\documentclass[aps,prl,reprint,superscriptaddress]{revtex4-1}

\usepackage{graphicx}
\usepackage{dcolumn}
\usepackage{amsmath}

\begin{document}

\title{Suppressed magnetic circular dichroism and valley-selective magneto-absorption due to the effective mass anisotropy in bismuth}

\author{Pieter J. de Visser}
\email{p.j.devisser@tudelft.nl}
\affiliation{Department of Quantum Matter Physics, University of Geneva, Geneva 1211, Switzerland}
\affiliation{Kavli Institute of NanoScience, Faculty of Applied Sciences, Delft University of Technology, Lorentzweg 1, 2628 CJ Delft, The Netherlands}

\author{Julien Levallois}
\email{Julien.Levallois@unige.ch}
\affiliation{Department of Quantum Matter Physics, University of Geneva, Geneva 1211, Switzerland}

\author{Micha\"{e}l K. Tran}
\affiliation{Department of Quantum Matter Physics, University of Geneva, Geneva 1211, Switzerland}

\author{Jean-Marie Poumirol}
\affiliation{Department of Quantum Matter Physics, University of Geneva, Geneva 1211, Switzerland}

\author{Ievgeniia O. Nedoliuk}
\affiliation{Department of Quantum Matter Physics, University of Geneva, Geneva 1211, Switzerland}

\author{J\'{e}r\'{e}mie Teyssier}
\affiliation{Department of Quantum Matter Physics, University of Geneva, Geneva 1211, Switzerland}

\author{Ctirad Uher}
\affiliation{Department of Physics, University of Michigan, Ann Arbor, Michigan 48109, USA}

\author{Dirk van der Marel}
\affiliation{Department of Quantum Matter Physics, University of Geneva, Geneva 1211, Switzerland}

\author{Alexey B. Kuzmenko}
\email{Alexey.Kuzmenko@unige.ch}
\affiliation{Department of Quantum Matter Physics, University of Geneva, Geneva 1211, Switzerland}

\date{\today}

\begin{abstract}

We have measured the far-infrared reflectivity and Kerr angle spectra on a high-quality crystal of pure semimetallic bismuth as a function of magnetic field, from which we extract the conductivity for left- and right handed circular polarisations. The high spectral resolution allows us to separate the intraband Landau level transitions for electrons and holes. The hole transition exhibits 100\% magnetic circular dichroism, it appears only for one polarisation as expected for a circular cyclotron orbit. However the dichroism for electron transitions is reduced to only $13\pm 1$\%, which is quantitatively explained by the large effective mass anisotropy of the electron pockets of the Fermi surface. This observation is a signature of the mismatch between the metric experienced by the photons and the electrons. It allows for a contactless measurement of the effective mass anisotropy and provides a direction towards valley polarised magneto-optical pumping with elliptically polarised light.
\end{abstract}

\maketitle

Circular dichroism, the property of materials to interact differently with left and right circularly polarised light is associated with symmetry breaking due to e.g. molecular chirality, spontaneous magnetisation or an external magnetic field. A charged particle in a magnetic field moves in a circular orbit perpendicular to that field. Radiation propagating parallel to the field is only absorbed by that particle for right \emph{or} left circular polarisation, leading to 100\% magnetic circular dichroism. In condensed matter, the behaviour of charge carriers in a lattice is described using an effective mass, which can be very different from the free-electron mass, but also strongly anisotropic \cite{Kittel}. In the latter case the cyclotron orbits become elliptical which, as we will show, strongly reduces the magnetic circular dichroism. The reduction of magnetic circular dichroism can thus be regarded as a manifestation of the mismatch between the metric experienced by photons (isotropic) and the electrons (anisotropic), which becomes apparent when they interact \cite{pwolff1964,fhaldane2011,byang2012}. As a consequence, elliptically polarised light with the same ellipticity as an electron pocket can cause a 100\% valley polarised magneto-optical absorption.

\begin{figure}
\includegraphics[width=0.99\columnwidth]{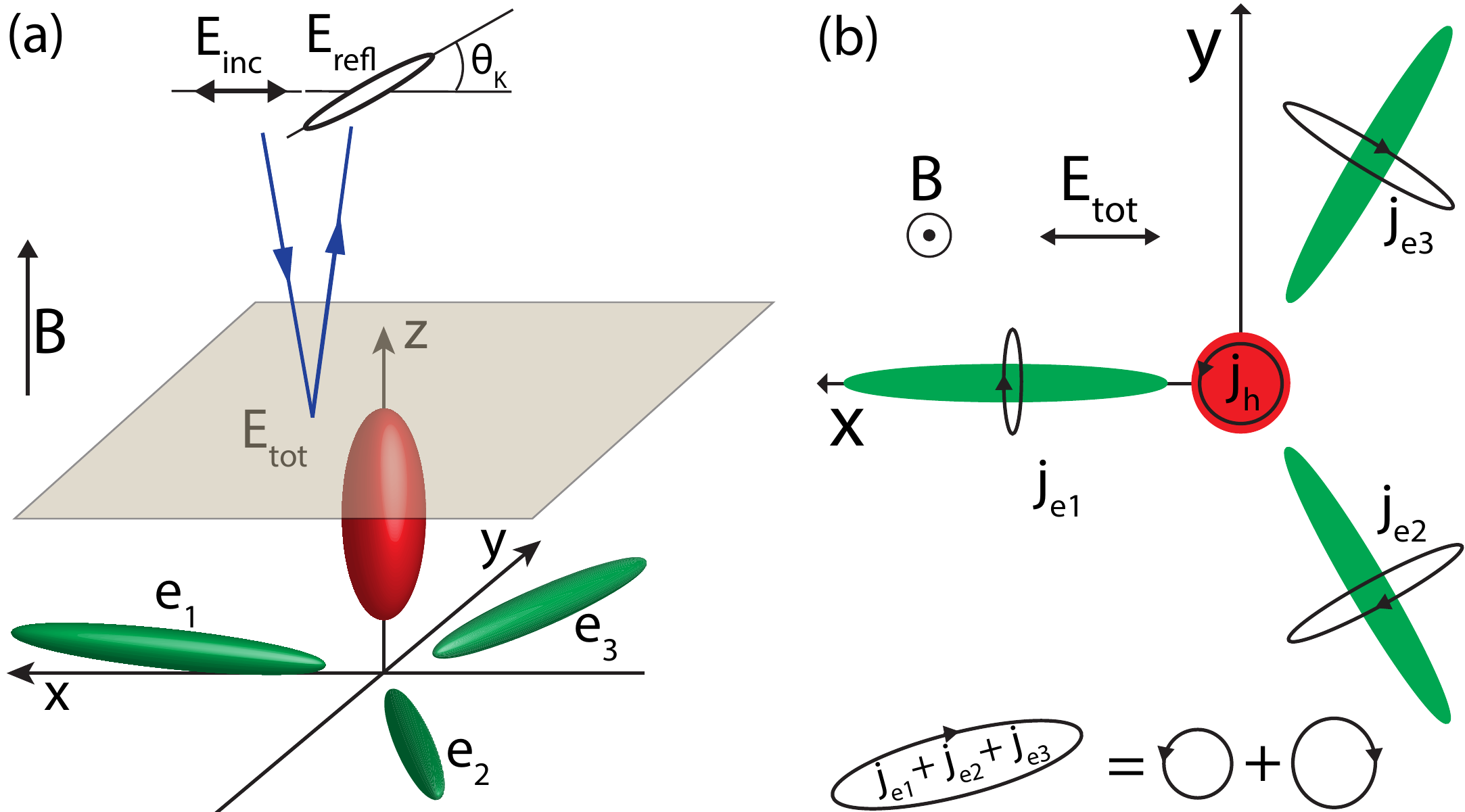}
\caption{\label{fig:schematic} (a) Schematic representation of the interaction of charge carriers in bismuth with light. The Fermi surface of bismuth is sketched in the interior of the sample, with a hole pocket (red) oriented along the trigonal ($z$) axis (inspired by Ref. \onlinecite{zzhu2011}). Three electron pockets (green) are oriented almost perpendicularly to the trigonal axis (with a 6$^{\circ}$ tilt), rotated by 120$^{\circ}$ with respect to each other. A magnetic field is applied along the $z$-axis. Radiation with linear polarisation is propagating almost parallel to the field. Close to the intraband resonances the light reflects with an elliptical polarisation and is rotated by the Kerr angle. (b) Top view on the trigonal plane. The effective electric field $E_{tot}$ in the crystal generates currents (both in real space) in the different pockets (in reciprocal space) according to $j=\sigma E_{tot}$. The current in the hole pocket results in complete magnetic circular dichroism. The sum of the current in the electron pockets shows strongly reduced dichroism due to the large effective mass anisotropy (bottom).}
\end{figure}

Bismuth is a canonical semimetal which possesses a rich electronic structure with strong spin-orbit interaction, low carrier density, long mean-free path and small cyclotron mass \cite{mcohen1960,lfalkovsky1968,vedelman1977,yfuseya2015}. Important phenomena such as the Seebeck effect \cite{tseebeck1823}, Nernst effect \cite{avettingshausen1886} and quantum oscillations \cite{lshubnikov1930,wdehaas1930} were discovered in bismuth. More recently a valley-ferromagnetic state \cite{lli2008}, a valley-nematic Fermi liquid state \cite{zzhu2011b}, a valley dependent density of states \cite{rkucher2014}, an avoided Lifshitz type semimetal-semiconductor transition \cite{narmitage2010}, and topological edge states \cite{idrozdov2014} have been revealed. The Fermi surface of bismuth is highly anisotropic (Fig. \ref{fig:schematic}a). A hole pocket is oriented along the trigonal axis, where the bands are almost parabolic. Three electron pockets are tilted 6$^\circ$ from the plane perpendicular to the trigonal axis and have a Dirac-like band dispersion. In this plane the dielectric function is isotropic. A magnetic field parallel to the trigonal axis combined with light propagating along the same axis therefore presents an ideal case to measure the dichroism in relation to the effective mass. For holes the mass is isotropic in the plane whereas for electrons it is strongly anisotropic (a factor $>$ 200 \cite{zzhu2011}), leading to a strong dichroism contrast between electron and hole transitions, as schematically shown in Fig. \ref{fig:schematic}b. Quite a few (far)infrared magneto-optical studies have been carried out to study band parameters, Landau level transitions and selection rules in Bi \cite{blax1960,wboyle1960,rbrown1963,jburgiel1965,gsmith1964,gbaraff1965,mmaltz1970,rblewitt1973,mvecchi1974,mvecchi1976,hverdun1976,hverdun1977}. However these were limited to selected wavelengths or, when spectroscopy was done, the polarisation dependence was not measured \cite{jomaggio1993,adelaforge2010}.

Here we present measurements of the far-infrared reflectivity and Kerr angle spectra on a high-quality crystal of pure bismuth as a function of magnetic field (0-7 T, applied along the trigonal axis), from which the conductivity for left- ($\sigma_-$) and right-handed ($\sigma_+$) circular polarisations (seen from the source) are derived. The high spectral resolution (1 cm$^{-1}$) enables distinguishing the intraband Landau level transitions for electrons and holes using their different magnetic field dependence. The observed hole transition shows maximum magnetic circular dichroism, it appears only for one polarisation, because the effective mass for holes is isotropic in this plane. However the magnetic circular dichroism, $A=(\sigma_+-\sigma_-)/(\sigma_++\sigma_-)$, for electron transitions is strongly reduced. The observed dichroism for the electron transitions of 0.13$\pm$0.01 agrees with calculations which include the strong effective mass anisotropy of the three electron pockets.

\begin{figure}
\includegraphics[width=0.99\columnwidth]{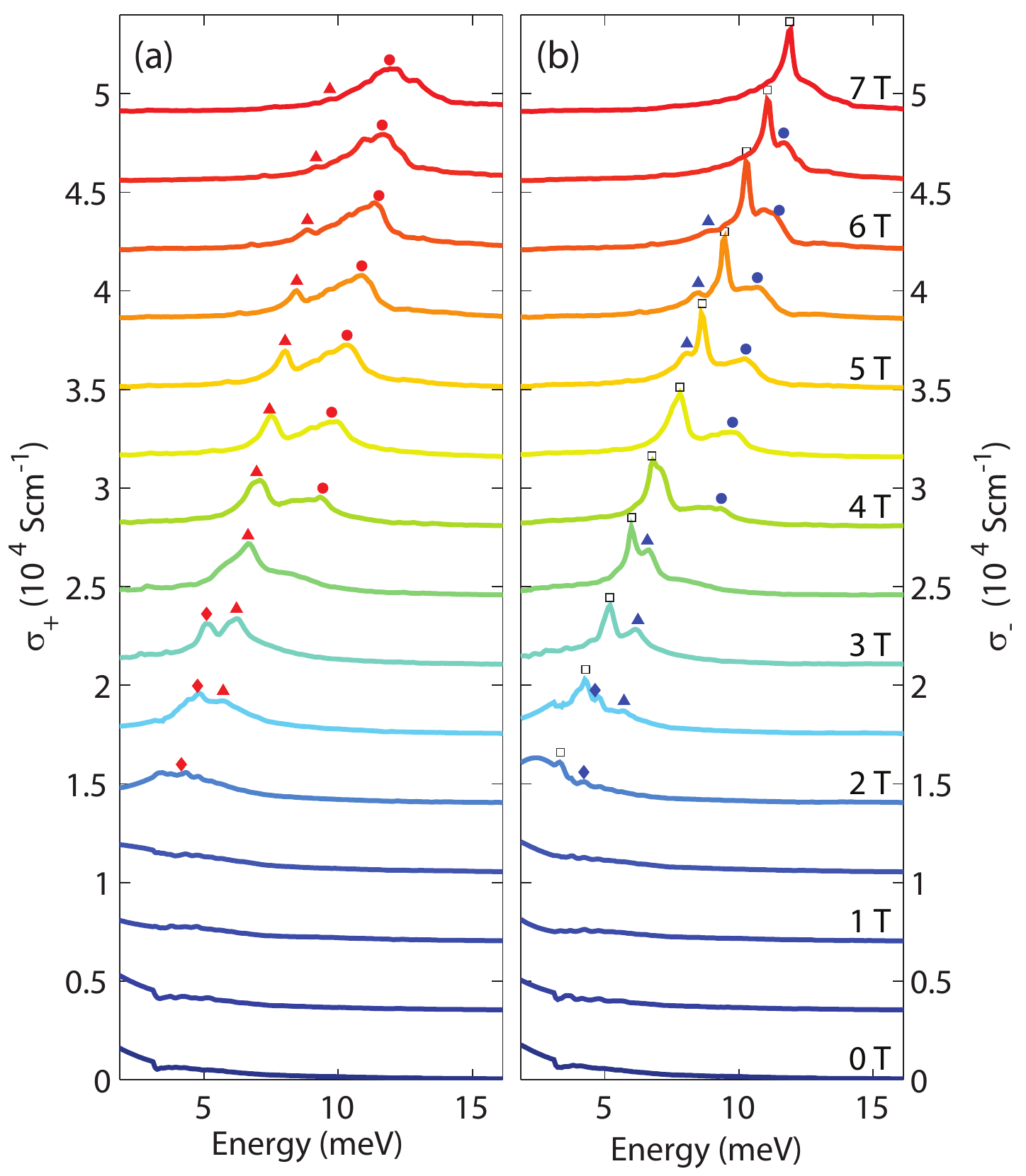}
\caption{\label{fig:sigmapm} (a,b) The real part of the measured conductivity for right and left circularly polarised light, $\sigma_+$ and $\sigma_-$, as a function of the photon energy. The magnetic field steps are 0.5 T and the same colour coding applies to both panels. Curves are displaced for clarity. The symbols at each peak correspond to Figure \ref{fig:energylevels}c,d.}
\end{figure}

The magneto-optical reflectivity and Kerr angle spectra were measured in a split-coil superconducting magnet attached to a Fourier transform spectrometer. The Bi crystal, with 99.9999\% purity, was cleaved in liquid nitrogen along the natural cleavage plane perpendicular to the trigonal axis. We performed spectroscopy on a freshly cleaved, untreated surface, which shows only atomic height variations over large parts of the surface (see Supplemental Material \cite{supinf}). The measurements were performed at 5 K with radiation at nearly normal angle incidence (8$^{\circ}$). The magnetic field was applied along the trigonal axis and thus almost parallel to the propagation of light (Faraday geometry).

The reflectivity $R(\omega,B)$ and Kerr angle $\theta_K(\omega,B)$ were measured in a single measurement run using two polarisers, one before and one after the sample. The absolute reflectivity was measured with a double reference method. At every magnetic field the spectrum on the sample and on a gold reference mirror was taken. The measurements were repeated after in-situ evaporation of a gold layer on the sample. The Kerr angle was measured using the fast protocol explained in Ref. \onlinecite{jlevallois2015}. More details on the experiment are given in the Supplemental Material \cite{supinf}, together with plots of the measured $R(\omega)$ and $\theta_K(\omega)$.

\begin{figure}
\includegraphics[width=0.99\columnwidth]{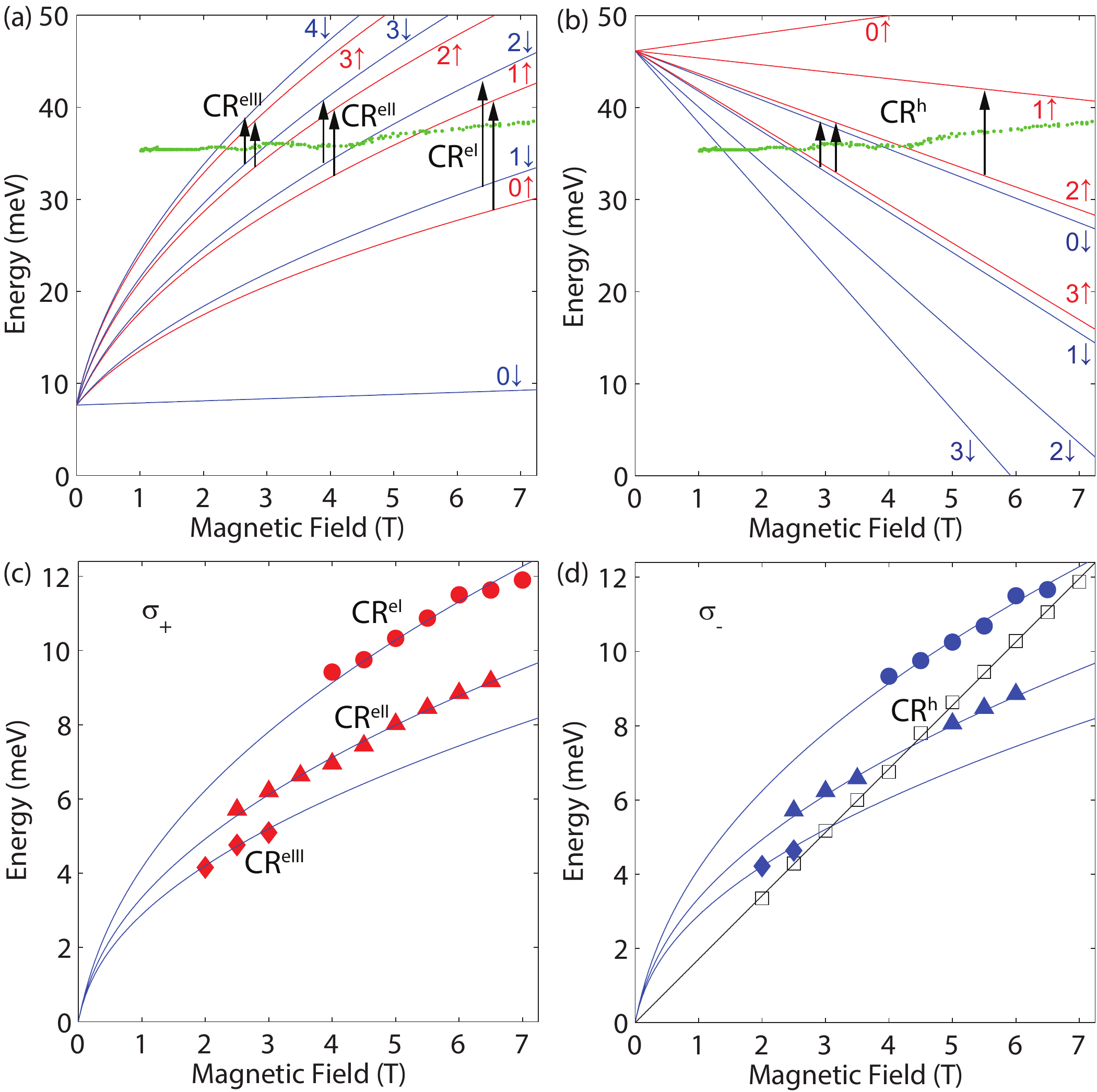}
\caption{\label{fig:energylevels} Energies of the Landau levels for electrons (a) and holes (b), labelled with their $n$ and spin. The Fermi level (green dots) is taken from Ref. \cite{zzhu2011}. The relevant transitions for this experiment are indicated as arrows. For electrons transitions for spin up from level $n$ have the same energy as transitions for spin down from level $n+1$. For holes all transitions have the same energy. (c,d) Measured energies of the Landau level transitions as a function of magnetic field for $\sigma_+$ and $\sigma_-$. The different symbols mark experimental points of the different series of transitions. The lines represent the theoretical field dependence as given by Equations \ref{eq:holetransition} (holes) and \ref{eq:electrontransition} (electrons). }
\end{figure}

We employ a generalised magneto-optical Kramers-Kronig analysis (MOKKA) to calculate $\sigma_-$ and $\sigma_+$ from the measurements \cite{jlevallois2015}. $R(\omega)$ and $\theta_K(\omega)$ are initially fitted using a sum of Lorentzians, which relies on complementary ellipsometry data for the high frequencies (not shown). Subsequently two independent Kramers-Kronig-consistent variational functions are added to the initial fit, and adjusted to simultaneously reproduce $R(\omega)$ and $\theta_K(\omega)$, resulting in model-independent $\sigma_+(\omega)$ and $\sigma_-(\omega)$, which are shown in Fig. \ref{fig:sigmapm}a and b for the energy range where intraband Landau-level transitions occur. The transition energies, which are marked on the peak positions in Fig. \ref{fig:sigmapm}a,b, are plotted with the same symbols in Fig. \ref{fig:energylevels}c,d as a function of magnetic field. Before we can address the magnetic circular dichroism, we first need to identify the various transitions that occur.

The hole band of the Brillouin zone in Bi is usually considered to be parabolic and therefore the hole Landau levels are equidistant in energy. Due to the influence of outside bands and a strong spin-orbit coupling, a Zeeman-like term needs to be added, resulting in the Landau level energies for the hole band $E_h$ as a function of magnetic field $B$ at $k_z=0$ given by \cite{zzhu2011}
\begin{equation}
E_h(n,s,B) = E_0+\Delta-\left( n+\frac{1}{2} \right)\frac{\hbar eB}{M_c} + sG\mu_BB,
\label{eq:holetransition}
\end{equation}
where $n=0,1,...$ is the Landau level index, $s=\pm\frac{1}{2}$ the spin quantum number, $E_0$ is the energy separation between the top of the hole valence band and the bottom of the electron conduction band ($E_0 = 38.5$ meV \cite{gsmith1964}), $e$ the elementary charge, $M_c$ the cyclotron mass, $\hbar$ Planck's reduced constant, $\mu_B$ the Bohr magneton and $G$ the g-factor for holes. The resulting energy levels are shown in Fig. \ref{fig:energylevels}b. The transition energy between levels with different $n$ is thus linear in $B$. 

For the electron pockets the gap separating the valence and conduction band is small ($2\Delta=15.3$ meV \cite{zzhu2011}) giving rise to a strongly relativistic Dirac-like band structure. Although the detailed electronic structure of Bi is complicated, most of the magnetotransport data can be described using a two-band Dirac Hamiltonian, complemented with a Zeeman term which approximately takes into account outside band effects \cite{mmaltz1970}. This extended Dirac Hamiltonian leads to the following expression for the Landau levels in the conduction band \cite{mmaltz1970,zzhu2011}:
\begin{equation}
E_e(n,s,B) = \sqrt{\Delta^2+2\Delta\left(n+\frac{1}{2}+s\right)\frac{\hbar eB}{m_c}}+sg'\mu_BB,
\label{eq:electrontransition}
\end{equation}
with $m_c$ the cyclotron mass and $g'$ the electron g-factor. The energy levels are shown in Fig. \ref{fig:energylevels}a. Spin-conserving transitions between the electron Landau levels have a quasi-square-root dependence on $B$. In writing the transition energies Eqs. \ref{eq:holetransition} and \ref{eq:electrontransition} we assumed that $k_z\approx 0$, since the Fermi pockets only span a small piece of k-space around the Fermi energy.

We can now identify the strong transition in $\sigma_-$ with the linear field dependence (Fig. \ref{fig:energylevels}d, open squares) as a hole transition with $\Delta n=-1$ and $\Delta s=0$. We fit the theoretical field dependence to the data (black line in Fig. \ref{fig:energylevels}d), which results in $M_c = 0.0677 \pm 0.0002 m_0$, in excellent agreement with literature \cite{zzhu2011}. Here $m_0$ is the free electron mass.

\begin{figure}
\includegraphics[width=0.99\columnwidth]{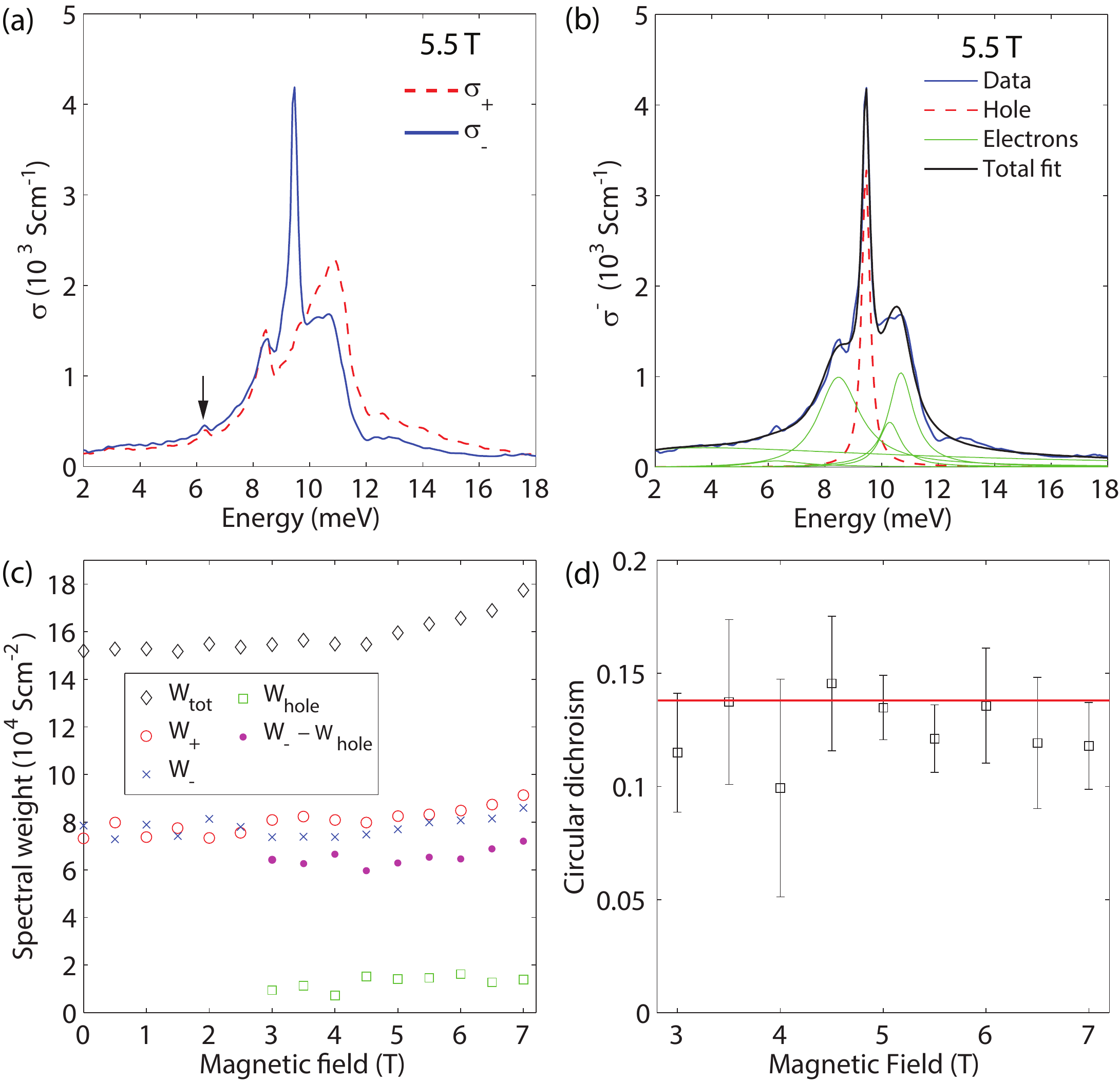}
\caption{\label{fig:spectral_weight} (a) Comparison of $\sigma_+$ and $\sigma_-$ as a function of energy measured at 5.5 T, showing the large dichroism for the hole transition together with the strongly reduced dichroism for the electron transitions. The tiny peak indicated with an arrow is discussed in the text. (b) Example of how the spectral weight for the hole transitions is derived for the $\sigma_-$ spectrum at 5.5 T. We use a Lorentzian at each electron transition and a Drude term to account for the low frequency spectral weight. (c) The spectral weight of the intraband transitions, integrated over energies 0-25 meV as a function of magnetic field. The symbols denote the total spectral weight in $\sigma_+ + \sigma_-$ (diamonds), in $\sigma_+$ (circles), in $\sigma_-$ (crosses), in the hole transition only (squares) and in $\sigma_-$ when the hole spectral weight is subtracted (dots). (d) The magnetic circular dichroism ratio for the electron transitions at the magnetic fields where the hole peak can be subtracted. The line is the theoretical value based on the effective mass anisotropy of the electron pockets. Error bars are explained in the text. }
\end{figure}

The transitions marked in Fig. \ref{fig:energylevels}c,d as filled symbols correspond to three different electron transitions with $\Delta n=1, \Delta s=0$. Fits through the data points using Eq. \ref{eq:electrontransition}, shown as blue lines in Fig. \ref{fig:energylevels}c and d, give $m_c = 0.0135 \pm 0.0001 m_0$ based on $2\Delta = 15.3$ meV \cite{zzhu2011}\footnote{Based on only this data, $\Delta$ and $m_c$ cannot be fitted independently.}. We note that the hole transition occurs sharper in the spectra (Fig. \ref{fig:sigmapm}) than the electron transitions because only the electron transitions are affected by a small spread in $k_z$ \cite{zzhu2011}. The arrow in Fig. \ref{fig:spectral_weight}a marks an unidentified peak, which has negligible spectral weight and is discussed in the Supplemental Material \cite{supinf}.

We have seen that the hole transition is only visible in $\sigma_-$ as expected for a circular cyclotron orbit. It is striking that all the observed electron transitions in $\sigma_+$ also show up in $\sigma_-$. Fig. \ref{fig:spectral_weight}a compares the measured $\sigma_+(\omega)$ and $\sigma_-(\omega)$ at 5.5 T. Apart from the strong dichroism at the hole peak, we observe that the spectral weight of the electron transitions is almost the same in $\sigma_+(\omega)$ and $\sigma_-(\omega)$. As we will now show, this is due to the strong effective mass anisotropy of the electron pockets.

To describe the anisotropy we first introduce a new coordinate system $\tilde{x}$ and $\tilde{y}$, in which the Fermi surface is isotropic, by using scaling coefficient $\alpha$.
\begin{equation}
x\rightarrow \tilde{x}= \alpha^{-1/2}x\mbox{ , } y\rightarrow \tilde{y}= \alpha^{1/2} y.
\label{eq:MetricTransformationX}
\end{equation}
Formally this is the same as transforming the effective mass and retaining the original metric:
\begin{equation}
m_{x}\rightarrow \tilde{m}_{x}=\alpha m_{x}\mbox{ , } m_{y}\rightarrow \tilde{m}_{y}=\alpha^{-1}m_{y}.
\label{eq:MetricTransformationM}
\end{equation}
The effective mass anisotropy is thus formally equivalent to a uniaxial stretching of the spatial metric \cite{pwolff1964,jquinn1964,byang2012}. The metric transformation conserves both the volume of the pocket and the cyclotron mass.

We proceed by considering the projection onto the trigonal plane of one electron pocket with its long axis parallel to the $x$-axis (Fig. \ref{fig:schematic}b). The conductivity from this pocket $\hat{\sigma}_{e1}$ can be written as
\begin{equation}
\hat{\sigma}_{e1} = \begin{pmatrix} \tilde{\sigma}_{xx} & \tilde{\sigma}_{xy} \\ -\tilde{\sigma}_{xy} & \tilde{\sigma}_{yy} \end{pmatrix} = \begin{pmatrix} \alpha^{-1}s_{xx} & s_{xy} \\ -s_{xy} & \alpha s_{xx} \end{pmatrix},
\label{eq:sigmae1}
\end{equation}
where $s$ and $\tilde{\sigma}$ are the conductivities for respectively an isotropic and anisotropic pocket. The derivation of Eq. \ref{eq:sigmae1}, for a general Hamiltonian, is given in the Supplemental Material \cite{supinf}. The magnetic circular dichroism for the isotropic case is thus given by $A^s=(s_+-s_-)/(s_++s_-)$, with $s_{\pm} = s_{xx}\pm is_{xy}$. Using Eq. \ref{eq:sigmae1}, the total electron conductivity is now obtained by adding the two remaining pockets which are rotated using rotation matrix $\hat{R}$ by $\pm 120^{\circ}$: $\hat{\sigma} = \hat{\sigma}_{e1} + \hat{R}\hat{\sigma}_{e1}\hat{R}^{-1} + \hat{R}^{-1}\hat{\sigma}_{e1}\hat{R}$. In the circular basis, ${\sigma}_{\pm}={\sigma}_{xx}\pm i{\sigma}_{xy}$, which leads to
\begin{equation}
{\sigma}_{\pm} = 3\left[ \left(\frac{\alpha^{-1/2}\pm \alpha^{1/2}}{2}\right)^2s_{+} + \left(\frac{\alpha^{-1/2}\mp \alpha^{1/2}}{2}\right)^2s_{-} \right].
\label{eq:sigmapmtot}
\end{equation}
The total dichroism $A$ can now be related to the effective mass anisotropy and is given by
\begin{equation}
A = \frac{{\sigma}_+-{\sigma}_-}{{\sigma}_++{\sigma}_-} = \frac{2}{\alpha^{-1}+\alpha}\frac{s_+-s_-}{s_++s_-} = \frac{2}{\alpha^{-1}+\alpha}A^s,
\label{eq:dichroismmass}
\end{equation}
regardless of the form of $A^s$ and valid for a general Hamiltonian and thus for parabolic and Dirac bands. For a circularly symmetric Fermi surface, $A = A^s$ and the dichroism is just that given by the transitions involved. In bismuth, we expect 100\% dichroism for the hole transitions (Fig. \ref{fig:schematic}b), which we experimentally observe in Figures \ref{fig:sigmapm} and \ref{fig:spectral_weight}a. For the electron transitions we also have $A^s = 1$, but the electron pockets are strongly elongated with $m_x = 0.257 m_0$ and $m_y = 0.00124 m_0$ (\cite{zzhu2011}, which takes into account the 6$^\circ$ tilt) and thus $\alpha = \sqrt{m_x/m_y} = 14.4$. Therefore we expect a strongly reduced dichroism, $A = 0.138$, for the electron transitions.

To derive the magnetic circular dichroism for electrons from the experimental data, we first need to subtract the spectral weight of the hole transition from $\sigma_-$. To that end we perform a fit using four Lorentzians and a small Drude term as shown in Fig. \ref{fig:spectral_weight}b. Alternatively the hole weight is calculated from the knowledge of the shape of $\sigma_+$. The plotted hole weight is an average of both methods \cite{supinf}. Figure \ref{fig:spectral_weight}c shows the spectral weight, $W=\int_0^{\mathrm{25~meV}} \sigma(\omega)d\omega$ (the intraband region), for $\sigma_+$, for $\sigma_-$, for $\sigma_+ + \sigma_-$, for the hole peak and for $\sigma_-$ with the hole peak subtracted. The increase of the total spectral weight above 4.5 T is related to the increase in carrier density at higher fields \cite{zzhu2011}. We can now derive the experimental dichroism for the electron pockets, which is shown in Fig. \ref{fig:spectral_weight}d and is the central result of this work. Error bars are estimated from the hole subtraction procedure \cite{supinf}. The line is the theoretical expectation $A = 0.138$, which agrees well with the measurements. The weighted average over the magnetic fields is $A=0.13\pm 0.01$. We can thus distinguish Landau level transitions that happen at the same time in an isotropic hole valley and in an anisotropic electron valley from each other. Moreover we can quantitatively relate the measured magnetic circular dichroism to the effective mass anisotropy.

Magnetic circular dichroism thus provides a contactless method to measure the effective mass anisotropy and could complement transport based techniques. To derive the dichroism, the anisotropy in the Hamiltonian has to be considered, in contrast to transport measurements where an isotropic mass model suffices (once the direction of the magnetic field is fixed) \cite{zzhu2011,byang2012}. The high mass anisotropy presents light which enters the crystal from vacuum with a large metric-mismatch. The photons can therefore still probe the mass anisotropy, using the reduced circular dichroism. Intriguingly, electronic interactions can also strongly influence the effective metric of a material \cite{bhalperin1993,fhaldane2011,byang2012}, which also plays a role in topological insulators, especially if they already have a band-mass anisotropy \cite{dhsieh2008,hzhang2009,pdiawa2012,mneupane2014,yzhao2015}. 

\begin{figure}
\includegraphics[width=0.99\columnwidth]{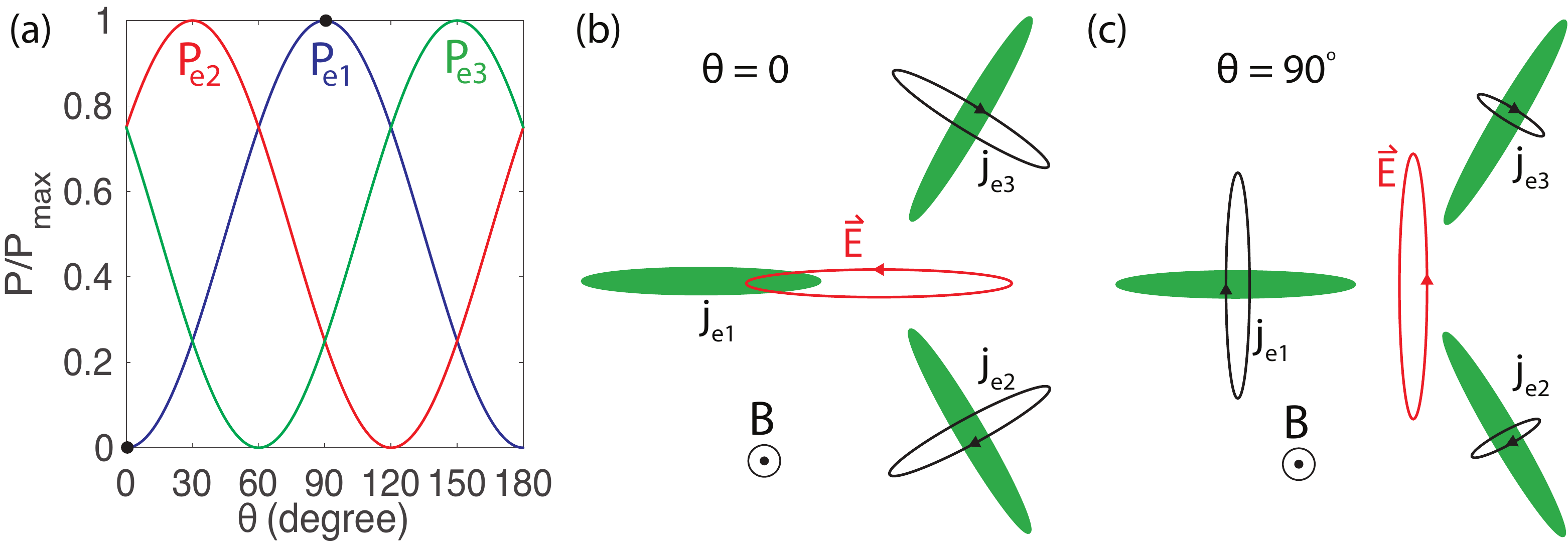}
\caption{\label{fig:valley_pol} (a) Relative power absorbed in each electron pocket for elliptically polarised light as a function of the azimuth angle of the polarisation ellipse $\theta$. (b) and (c) Schematics of the valley-polarised magneto-absorption for  $\theta=0$ and $\theta=90^\circ$ respectively.}
\end{figure}

The above measurements and theory suggest that applying elliptically polarised light would lead to a highly valley-polarised magneto-absorption. If we consider light with an ellipticity matching the cyclotron orbits with opposite handedness, the absorbed power per unit volume in pocket e1 is given by \cite{supinf}
\begin{equation}
P_{e1}(\theta) =P_{e1,max}\sin^{2}\theta,
\end{equation}
with $\theta$ the angle between the long-axis of the polarisation ellipse and the $x$-axis. Consequently $P_{e2}(\theta) = P_{e1}(\theta+60^{\circ})$ and $P_{e3}(\theta) = P_{e1}(\theta-60^{\circ})$, which are plotted in Fig. \ref{fig:valley_pol}a. A 100\% valley polarisation (ie one valley does not absorb light) can thus be obtained at three angles $\theta = 0,60,120^\circ$ (Fig. \ref{fig:valley_pol}). This is a totally different approach to valleytronic applications in bismuth, as compared to previous suggestions where the magnetic field is rotated in the xy plane \cite{apopescu2012,zzhu2011b}. It provides another perspective to valleytronics in dichalcogenides, in which circularly polarised light is normally used \cite{kbehnia2012,hzeng2012,kmak2012,tcao2012}, and to the proposed valley polarisation in topological insulators using elliptical polarisation \cite{mezawa2014}.

\begin{acknowledgments}
We acknowledge B. Fauqu\'{e}, K. Behnia, G.N. Kozhemyakin and T. Giamarchi for discussions and M. Brandt, S. Zanos, D. Chablaix, D. Stricker and I. Gaponenko for technical assistance. This research was supported by the Swiss National Science Foundation (grant 200020-156615) and by the EU Graphene Flagship (contract CNECT-ICT-604391). P.J. de V. acknowledges support from a Niels Stensen Fellowship.
\end{acknowledgments}

P.J. de V. and J. L. contributed equally to this work.

%

\clearpage
\onecolumngrid
\setcounter{figure}{0}
\setcounter{page}{1}
\setcounter{equation}{0}

\renewcommand{\thefigure}{S\arabic{figure}}

\section*{Supplementary material for: Suppressed magnetic circular dichroism and valley-selective magneto-absorption due to the effective mass anisotropy in bismuth} 

\subsection*{Pieter J. de Visser, Julien Levallois, Micha\"{e}l K. Tran, Jean-Marie Poumirol, Ievgeniia O. Nedoliuk, J\'{e}r\'{e}mie Teyssier, Ctirad Uher, Dirk van der Marel, Alexey B. Kuzmenko}

\maketitle

\section{Measurement of reflectivity and Kerr angle}
\begin{figure}[h]

\includegraphics[width=0.99\columnwidth]{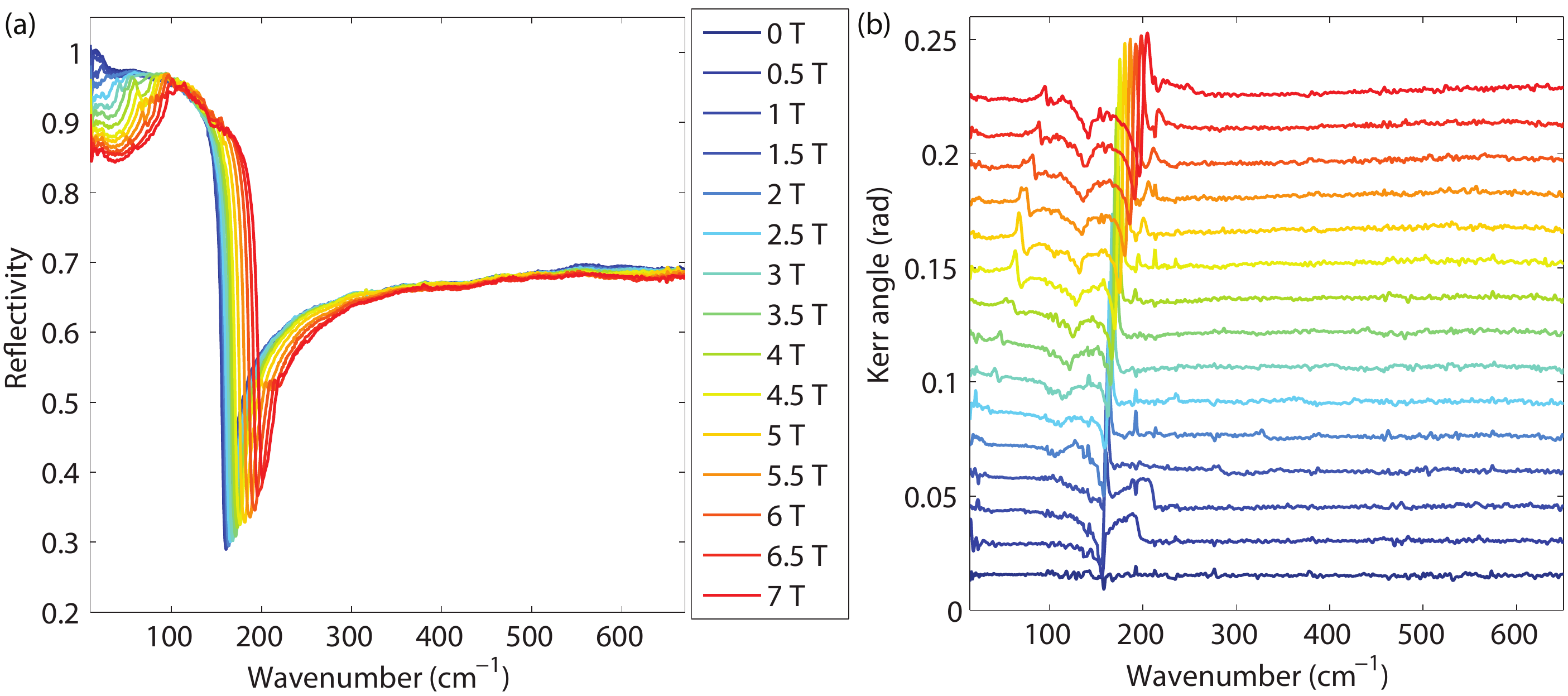}
\caption{\label{figS:reflkerr} Reflectivity (a) and Kerr angle (b) measured as a function of frequency for the magnetic fields indicated in the legend. The Kerr angle curves are shifted by 0.015 for clarity. The small spike that shows up at 193 cm$^{-1}$ for some fields in the Kerr angle is due to microphonic noise.}

\end{figure}

The reflectivity and Kerr angle spectra were measured in a single measurement run using two polarisers, one before and one after the sample. For the lowest energies (2-12 meV) we have used a mercury source and a Si beamsplitter and for the higher energies (12-87 meV) a Globar source with a Ge-coated mylar beamsplitter. Both ranges were measured with He cooled bolometers with a resolution of 1 cm$^{-1}$ and 2 cm$^{-1}$ respectively, using a Bruker V70 spectrometer. The absolute reflectivity, $R(\omega,B)$, was measured with parallel polarisers, using a double reference method: $R(\omega,B) = (I_s(\omega,B)/I_{r,s}(\omega,B))/(I_g(\omega,B)/I_{r,g}(\omega,B))$. At every field, the intensity spectrum of the sample, $I_s(\omega,B)$, and a spectrum on a gold reference mirror $I_{r,s}(\omega,B)$ were taken, to compensate for temporal drifts and possible small magnetic field dependences of the alignment. The same set of measurements was performed after evaporating a gold layer on the sample, giving $I_g(\omega,B)$ on the gold-coated sample and $I_{r,g}(\omega,B)$ on the reference mirror.

The Kerr angle was measured using the so-called fast protocol, introduced and explained in detail in Ref. \onlinecite{jlevallois2015}, valid for small Kerr angles. It relies on measuring the ratio of the reflected intensity at positive and negative fields $\rho(\omega,B) = I(\omega,+B)/I(\omega,-B)$ at analyser angles of +45 and -45 degrees for both the sample and the mirror reference. The Kerr angle is then given by
\begin{equation}
\theta_K(\omega,B) \cong \frac{1}{8}\left(\frac{\rho_{+45}(\omega,B)}{\rho_{+45,r}(\omega,B)} - \frac{\rho_{-45}(\omega,B)}{\rho_{-45,r}(\omega,B)} \right).
\label{eq:Kerrmeas}
\end{equation}
Note that the evaporated gold reference is not needed for $\theta_K$.

Figure \ref{figS:reflkerr} shows the measured reflectivity and Kerr angle as a function of wavenumber for different magnetic fields. These experimental results are used to calculate the optical conductivity for left and right circularly polarised light as shown in Figure 2 of the main text.

\section{Derivation of the equation (5) from the main text}
Although the equation (5) is very general, we shall explicitly consider the two cases of parabolic and Dirac bands corresponding to the hole and electron Fermi pockets respectively. The hole Hamiltonian in zero magnetic field is well approximated by
\begin{equation}\label{Ham-h}
H_{h}=E_0+\Delta-\frac{\hbar^2(k_{x}^2+k_{y}^2)}{2M_{c}}-\frac{\hbar^2 k_{z}^2}{2M_{z}},
\end{equation}
\noindent where $M_{z}$ is the effective mass of holes along the $z$-axis, and the definition of other parameters is given in the main text. It was established long time ago \cite{pwolff1964} that the band structure near the electron pocket is in many respects similar to the famous four-component relativistic Dirac Hamiltonian describing electrons and positrons. Although the actual band structure of electrons in bismuth is highly anisotropic, it is often substituted by a simplified model\cite{zzhu2011}:
\begin{equation}\label{Ham-e}
H_{e}=\left[
\begin{matrix}
\Delta &  0 & i\hbar v_{z}k_{z} & i\hbar v(k_{x} - i k_{y})\\
0 &  \Delta & i\hbar v(k_{x} + i k_{y}) & -i\hbar v_{z}k_{z}\\
-i\hbar v_{z}k_{z} &  -i\hbar v(k_{x} - i k_{y}) & -\Delta & 0\\
-i\hbar v(k_{x} + i k_{y}) &  i\hbar v_{z}k_{z} & 0 & -\Delta\\
\end{matrix}\right],
\end{equation}
\noindent where $v=(\Delta/m_{c})^{1/2}$ and $v_{z}=(\Delta/m_{z})^{1/2}$ and $m_{z}$ is the effective mass of electrons along the $z$-axis. This Hamiltonian describes correctly the Landau level energies and therefore is suitable to fit the transport data , such as the Nernst effect \cite{zzhu2011}, and thermodynamic properties. However, it does not capture the strong anisotropy in the $xy$ plane and therefore cannot describe the effects reported in this article. Below we will show how to modify it properly.

For a magnetic field $B$ applied along the $z$ axis, we follow a standard procedure and  perform the substitution:
\begin{equation}\label{Kohn-iso}
k_x + ik_y \rightarrow \frac{l_B}{\sqrt{2}}a^{+}  \mbox{ , }
k_x - ik_y \rightarrow \frac{l_B}{\sqrt{2}}a^{-}
\end{equation}
 \noindent where ${l_B} = (\hbar/eB)^{1/2} $ is the magnetic length. The operators $a^{+}$ and $a^{-}$ act on harmonic oscillator-like wavefunctions $\psi_{n=0,1,2..}(B,x,y)$ obeying the ladder algebra:
\begin{equation}\label{Ladder}
a^{+}\psi_{n=0,1..} = (n+1)^{1/2}\psi_{n+1} \mbox{ , }
a^{-}\psi_{n=1,2..}  = n^{1/2}\psi_{n-1}  \mbox{ , } a^{-}\psi_{0} = 0.
\end{equation}
\noindent This results in:
\begin{equation}\label{Ham-h-B}
H_{h}(B)=E_0+\Delta-\hbar\omega_c \left(a^+a^-+\frac{1}{2}\right)-\frac{\hbar^2 k_{z}^2}{2M_{z}},
\end{equation}
\noindent and
\begin{equation}\label{Ham-e-B}
H_{e}(B)=\left[
\begin{matrix}
\Delta &  0 & i\hbar v_{z}k_{z} & i \frac{\sqrt{2}}{l_{B}}\hbar va^{-}\\
0 &  \Delta & i \frac{\sqrt{2}}{l_{B}}\hbar va^{+} & -i\hbar v_{z}k_{z}\\
-i\hbar v_{z}k_{z} &  -i \frac{\sqrt{2}}{l_{B}}\hbar va^{-} & -\Delta & 0\\
-i \frac{\sqrt{2}}{l_{B}}\hbar va^{+} &  i\hbar v_{z}k_{z} & 0 & -\Delta
\end{matrix}\right],
\end{equation}
\noindent  where we omitted for simplicity the Zeeman term, which does not affect the spin-conserving transitions. Diagonalisation of $H_{h}(B)$ and $H_{e}(B)$ provides the eigenvalues $E_{h}(n,k_z)$ and $E_{e}(n,k_z)$  with the corresponding wavefunctions $\Psi_{h}(n, x, y, k_{z})$ and $\Psi_{e}(n, x, y, k_{z})$, where $n$ is a Landau-level index. For the case of $k_z$ = 0, these eigenvalues are given in  Eqs. (1) and (2) of the main text (apart from the Zeeman component).

The complex magneto-optical conductivity tensor can be calculated using the Kubo formula:
\begin{equation}\label{Kubo}
s_{\beta\gamma}(\omega,B) = \frac{e^3 B}{\pi^2}\int d k_{z}\sum_{n,n'}\langle \Psi_{n}|V_{\beta}|\Psi_{n'}\rangle \langle \Psi_{n'}|V_{\gamma}^{*}|\Psi_{n}\rangle
\frac{f(E_{n})-f(E_{n'})}{E_{n'}-E_{n}} \frac{i}{\hbar\omega -E_{n'}+E_{n}+i\Gamma_{nn'}},
\end{equation}
\noindent where $\beta$ and $\gamma$ are the Cartesian indices ($x$, $y$ and $z$), $n$ and $n'$ are the Landau-level indices of the initial and final states respectively , $f(E)=\left(\exp{\frac{E-E_{F}}{k_{B}T}} + 1\right)^{-1}$ is the Fermi-Dirac distribution, $E_{F}$ is the Fermi energy, $T$  is temperature, $\Gamma_{nn'}$ is the transition broadening and $V_{\beta}=\frac{1}{\hbar}\frac{\partial H}{\partial k_{\beta}}$ is the velocity operator. Note that we use here $s$ here for the conductivity of an isotropic system instead of the conventional $\sigma$, for consistency with the main text, where we reserve $\sigma$ for the measured conductivity. For a system rotationally invariant in the $xy$ plane we obviously have: $s_{yy} = s_{xx}$ and $s_{yx} = -s_{xy}$.

Next we consider the effect of anisotropy. Although the actual band structure is quite complex, giving rise to a small tilting of the electron pockets, we shall restrict ourselves to the simplest case, where effective masses $m_{x}$ and $m_{y}$ along the $x$ and $y$ axes are different. The effective cyclotron mass then becomes $m_{c} = (m_{x}m_{y})^{1/2}$. It was conjectured \cite{byang2012} that such mass anisotropy can be formally removed ('gauged out')  by a proper spatial-metric transformation:
\begin{equation}\label{metric}
\tilde{x}=\alpha^{-1/2} x \mbox{ , }
\tilde{y}=\alpha^{1/2} y,
\end{equation}
\noindent or equivalently
\begin{equation}\label{metric2}
\tilde{k}_x=\alpha^{1/2} k_x \mbox{ , }
\tilde{k}_y=\alpha^{-1/2} k_y
\end{equation}
\noindent where $\alpha=(m_{x}/m_{y})^{1/2}$. In other words, the mass anisotropy is formally equivalent to a uniaxial stretching or squeezing of the spatial metric, in analogy with basic principles of general relativity. It is important that this deformation conserves not only the cyclotron mass but also the Fermi volume, the doping and the density of states. Therefore it does not influence most transport and thermodynamic properties. The transformation (\ref{metric}) maps an anisotropic model to the isotropic one. Indeed, in this case it is sufficient to redefine the a-operators:
\begin{equation}\label{Kohn-Aniso}
\tilde{k}_x + i\tilde{k}_y \rightarrow \frac{l_B}{\sqrt{2}}\tilde{a}^{+} \mbox{ , }
\tilde{k}_x + i\tilde{k}_y \rightarrow \frac{l_B}{\sqrt{2}}\tilde{a}^{-},
\end{equation}
\noindent which yields a Hamiltonian identical to (\ref{Ham-e-B}) but expressed in terms of $\tilde{a}^{\pm}$ instead of $a^{\pm}$. Therefore the Landau-level energies and wave-functions of an anisotropic Hamiltonian are easily expressed via the ones of the isotropic one:
\begin{eqnarray}\label{EBAnis}
\tilde{E}_{n}(B, k_{z})&=&E_{n}\left(B, k_{z}\right) \\
\tilde{\Psi}_{n}(B, x, y, k_{z})&=&\Psi_{n}\left(B, \alpha^{-1/2}x, \alpha^{1/2} y, k_{z}\right).
\end{eqnarray}
Our goal is to express the optical conductivity of a deformed system, $\tilde{\sigma}(\omega)$ in terms of the one of the isotropic one $s(\omega)$. If we take into account that the Landau level energies and the Fermi energy are not affected, while the velocity matrix elements are scaled as $\tilde{V}_{x}=\alpha^{-1/2} V_{x}$ and $\tilde{V}_{y}=\alpha^{1/2} V_{y}$, then, using the Kubo formula (\ref{Kubo}), we arrive straightforwardly at the equation (5) from the main text:
\begin{equation}\label{Eq5}
\begin{pmatrix}
\tilde{\sigma}_{xx} & \tilde{\sigma}_{xy} \\
-\tilde{\sigma}_{xy} & \tilde{\sigma}_{yy}
\end{pmatrix} =
\begin{pmatrix}
\alpha^{-1}s_{xx} & s_{xy} \\
-s_{xy} & \alpha s_{xx}
\end{pmatrix}.
\end{equation}
\noindent It is worth mentioning that in deriving this result we only used the fact that the original Hamiltonian is rotationally invariant, but did not use its specific form, i.e. equations (\ref{Ham-h-B}) or (\ref{Ham-e-B}). Therefore, the formula (5) from the main text is generally valid for any uniaxially deformed isotropic band structure.

\section{Valley-selective magneto-absorption for elliptical polarisation}

{A general elliptical polarization can be described by the Jones vector (in the $xy$ plane):
\begin{equation}\label{JonesVector}
\vec{E}=\frac{E_{0}}{\sqrt{1+\eta^2}}
\begin{pmatrix}
\cos\theta+i\eta\sin\theta \\
\sin\theta-i\eta\cos\theta
\end{pmatrix},
\end{equation}
\noindent where $\eta$ is the ellipticity (the ratio between the short and the long axes of the polarisation ellipse) and $\theta$ is the azimuth angle (the angle between the long axis and the $x$-axis). Note that $-1\leq\eta\leq1$ and is defined positive/negative for clockwise/counterclockwise rotations respectively. $\eta = \pm 1$ correspond to the two circular polarisations and $\eta = 0$ refers to a linear polarisation. The normalisation is chosen such that the radiation power does not depend on the ellipticity.

We shall assume that the total electric field in a certain location inside the sample is given by Equation (\ref{JonesVector}). The power (per unit volume) absorbed by electrons in the pocket e1 is given by:
\begin{equation}\label{Power}
P_{e1}=\Re(\vec{j}_{e1}^{*}\vec{E})=\Re(\vec{E}^{*}\hat{\sigma}_{e1}\vec{E}),
\end{equation}
\noindent where $\hat{\sigma}_{e1}$ is the optical conductivity tensor and $\vec{j}_{e1}$ is the current induced in this pocket. In the remaining two pockets e2 and e3 we have accordingly $P_{e2}=\Re(\vec{E}^{*}\hat{\sigma}_{e2}\vec{E})$ and $P_{e3}=\Re(\vec{E}^{*}\hat{\sigma}_{e3}\vec{E})$, where $\hat{\sigma}_{e2}=\hat{R}\hat{\sigma}_{e1}\hat{R}^{-1}$ and $\hat{\sigma}_{e3}=\hat{R}^{-1}\hat{\sigma}_{e1}\hat{R}$ ($\hat{R}$ is the rotation matrix by 120$^{\circ}$).

For intraband n-type Landau-level transitions the selection rule dictates that $s_{-} = 0$, which means that $s_{xx} = s_{+}/2$, $s_{xy} = -is_{+}/2$ in the Equation (5) from the main text:
\begin{equation}\label{SigmaMatrix}
\hat{\sigma}_{e1}=\frac{s_{+}}{2}
\begin{pmatrix}
\alpha^{-1} & -i \\
i & \alpha
\end{pmatrix},
\end{equation}
\noindent where we assume that $\alpha>1$ (in bismuth it is 14.4).

By substituting Equations (\ref{JonesVector}) and (\ref{SigmaMatrix}) into Equation (\ref{Power}), we get:
\begin{equation}\label{PowerThetaGen}
P_{e1}(\eta,\theta) = \frac{\Re (s_{+})E_{0}^{2}}{2}\left[\frac{\alpha^{-1}+\alpha}{2}+\frac{2}{\eta^{-1}+\eta}-\frac{\alpha^{-1}-\alpha}{2}\cdot\frac{\eta^{-1}-\eta}{\eta^{-1}+\eta}\cos2\theta\right].
\end{equation}
\noindent and correspondingly $P_{e2}(\eta,\theta) = P_{e1}(\eta,\theta+60^{\circ})$ and $P_{e3}(\eta,\theta) = P_{e1}(\eta,\theta-60^{\circ})$. One can see that for a given valley the absorbed power is modulated by the azimuth angle only if $\alpha > 1$ (anisotropic mass) and $\eta \neq\pm 1$ (non-circular polarisation). At the same time, the total power $P_{tot} = P_{e1} +P_{e2} +P_{e3}$ is azimuth-independent for any light ellipticity, which is expected for an optically isotropic system, such as bismuth in the $xy$-plane.

If the ellipticity of the electromagnetic wave is set to match the ellipticity of the Fermi pocket, and the rotation direction is opposite to the cyclotron motion of the electrons ($\eta = \alpha^{-1})$ then Equation (\ref{PowerThetaGen}) reduces to
\begin{equation}\label{PowerTheta}
P_{e1}(\theta) = \frac{\Re (s_{+})E_{0}^{2}}{2}\frac{(\alpha^{-1}-\alpha)^{2}}{\alpha^{-1}+\alpha}\sin^{2}\theta=P_{e1,max}\sin^{2}\theta.
\end{equation}
\noindent $P_{e1}$ becomes zero if $\theta = 0^{\circ}$, i.e. when the polarisation ellipse is orthogonal to the pocket e1, which is depicted in Fig. 5 in the main text together with plots of Eq. \ref{PowerTheta} for each pocket. Obviously, in the pockets e2 and e3 the absorption vanishes if $\theta$ is set to 120 and 60 degrees respectively.

\section{AFM of cleaved surface}
We have cleaved the Bi crystal in liquid nitrogen. The natural cleavage plane of a Bi crystal is perpendicular to the trigonal axis. To inspect the surface in detail, which looks very flat by the eye, we performed atomic force microscopy on it. At typical topographic image is shown in Figure \ref{figS:afm}a. A line profile is shown in Fig. \ref{figS:afm}b, showing height steps of approximately 0.4, 0.8 and 1.2 nm, which correspond to 1/3 cuts of the unit cell (which is 11.85 \r{A} high). By repeating several experiments on Bi crystals we have learned that the magneto-optics measurements on a pristine cleaved surface lead to the highest quality data (sharper transitions and more pronounced plasma features in reflectivity and Kerr angle).

\begin{figure}

\includegraphics[width=0.99\textwidth]{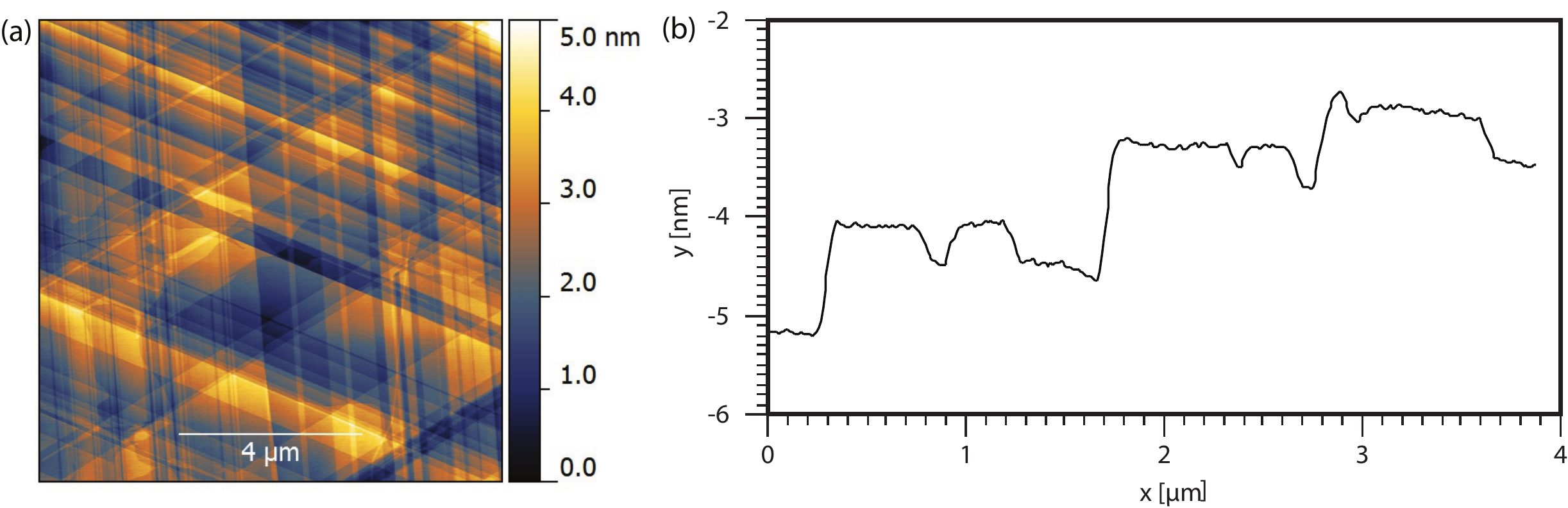}
\caption{\label{figS:afm} (a) False colour atomic force microscopy image of a cleaved Bi crystal surface. (b) A typical line profile which shows atomic height variations.}

\end{figure}

\section{Extraction of the hole spectral weight from $\sigma_-$ and its uncertainty}
To derive the circular dichroism for the electron pockets from the measured $\sigma_-$ and $\sigma_+$, we need to subtract the transition peak due to the holes from $\sigma_-$ as discussed in the main text. In there we have shown in Fig. 4b one method to estimate the hole weight, by fitting Lorentzians to the electron and hole transitions in $\sigma_-$ and subtracting the weight of the hole. Note that the goal of the Lorentzians fit is only to extract the spectral weight of the hole transition, not to perfectly describe the other spectral features. However, from a careful inspection of the measurements the peak due to the hole transition is not completely Lorentzian. From $\sigma_+$ we also observe that the main electron transition at the highest energy does not have a purely Lorentzian shape either. Both aspects lead to an uncertainty in the derived hole weight.

\begin{figure}

\includegraphics[width=0.49\textwidth]{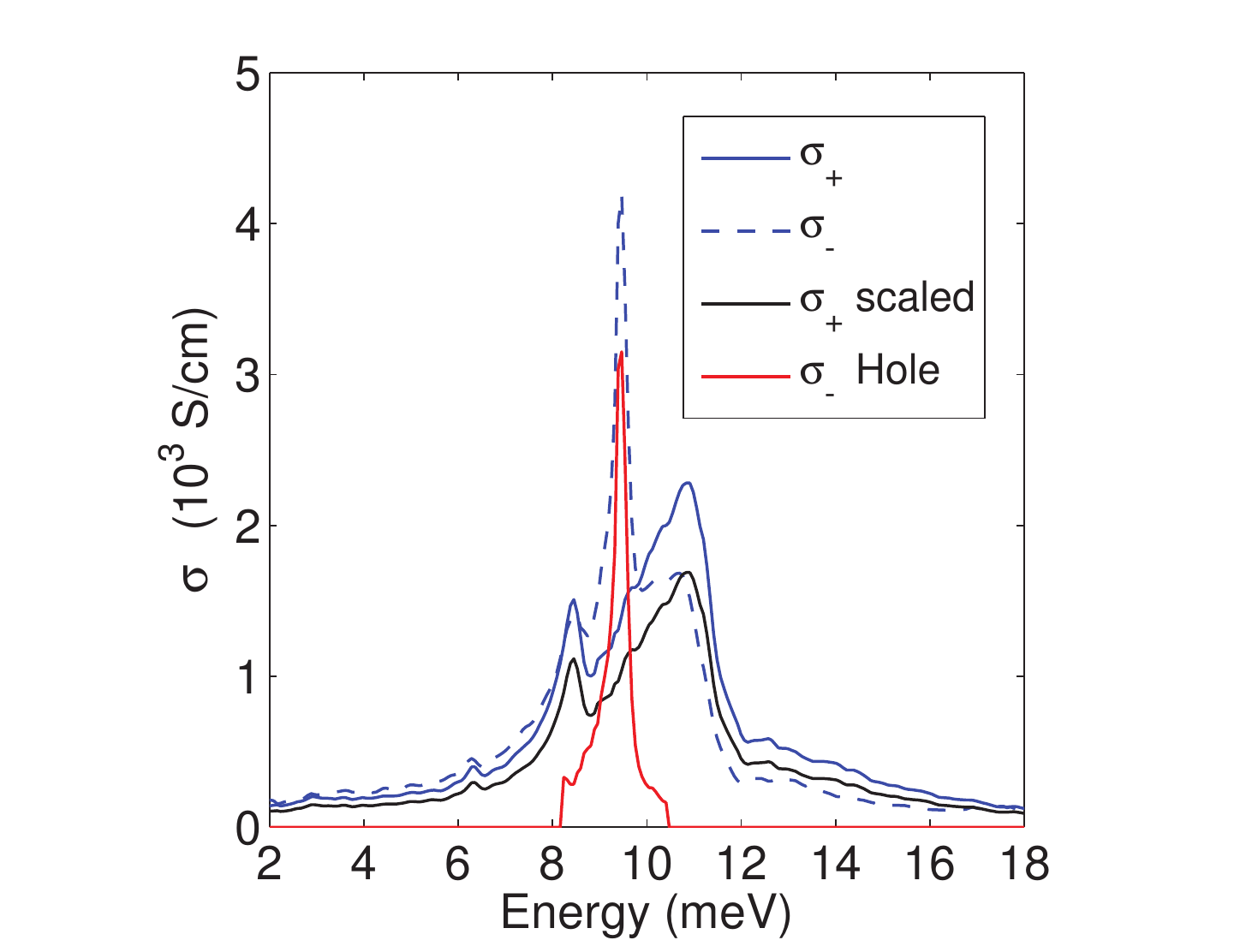}
\caption{\label{figS:methodhole2} Alternative way of deriving the hole spectral weight from the measured $\sigma_-$ and $\sigma_+$. $\sigma_+$, scaled to the height of the electron transitions in $\sigma_-$, is subtracted from the measured $\sigma_-$ to determine the weight of the hole transition in $\sigma_-$.}

\end{figure}

An alternative way to derive the weight in the hole peak is to use the knowledge that $\sigma_+$ only represents electron transitions. We can subsequently scale $\sigma_+$ to match the visible electron transitions in $\sigma_-$. We can then calculate $\sigma_{hole} = \sigma_--\sigma_{+,scaled}$ in the region where the hole peak dominates. This procedure is shown in Fig. \ref{figS:methodhole2}. This method has the advantage that it does not assume a particular shape for the electron and hole transitions. However, the shape of $\sigma_-$ and $\sigma_+$ due to the electron transitions is not exactly the same. Therefore we need to cut the integration boundaries at a finite point, which leaves an uncertainty on the spectral weight.

For the final dichroism as shown in Fig. 4d of the main text, we have taken the average dichroism obtained by the two methods described above. The uncertainty contains two parts. One part is taken to be the difference in dichroism between the two methods of hole subtraction. For the second part we estimate that each method by itself leads to an uncertainty in dichroism of 0.02, which adds to the average of the two methods an uncertainty $0.02/\sqrt{2}$. Together this leads to the error bars in Fig. 4d of the main text. The reported average dichroism $0.13\pm 0.01$ is calculated from a weighted average over the magnetic fields.

\section{Unidentified, low energy peak in the conductivity}
In the conductivity spectra in Fig. 2 of the main text one can identify a small additional peak at low energies. It is better visible in the spectra at 5.5 T as shown in Fig. 4a of the main text. It occurs in both $\sigma_-$ and $\sigma_+$ and its energy is plotted for the fields at which it is visible in Figure \ref{figS:smallpeak}. It cannot be identified in the same hole and electron picture since its field dependence does not correspond to the field dependence of Landau level, spin and combination resonances. The resonance shows up in both $\sigma_+$ and $\sigma_-$ and in this sense has characteristics similar to the electron-pockets near the L-point. However, the field dependence does not match the band structure of the bulk. A transition was observed before at the same energies (below the cyclotron frequencies) in an infrared transmission experiment \cite{jburgiel1965}, which was attributed in that paper to a combination transition (cyclotron and spin). However this explanation used a different frequency for cyclotron and spin transitions, which is not consistent with the extended two-band model which is generally accepted to describe the Landau level transitions in Bi and which we use in this work. We emphasise that the spectral weight of this feature is very small compared to the total spectral weight of the electron resonances and therefore does not influence the analysis of the circular dichroism. It is known that Bi exhibits surface states, which could also be a source of magnetic-field dependent transitions. However this experiment does not contain sufficient information to definitively ascribe this small peak to a surface state.

\begin{figure}
\includegraphics[width=0.49\textwidth]{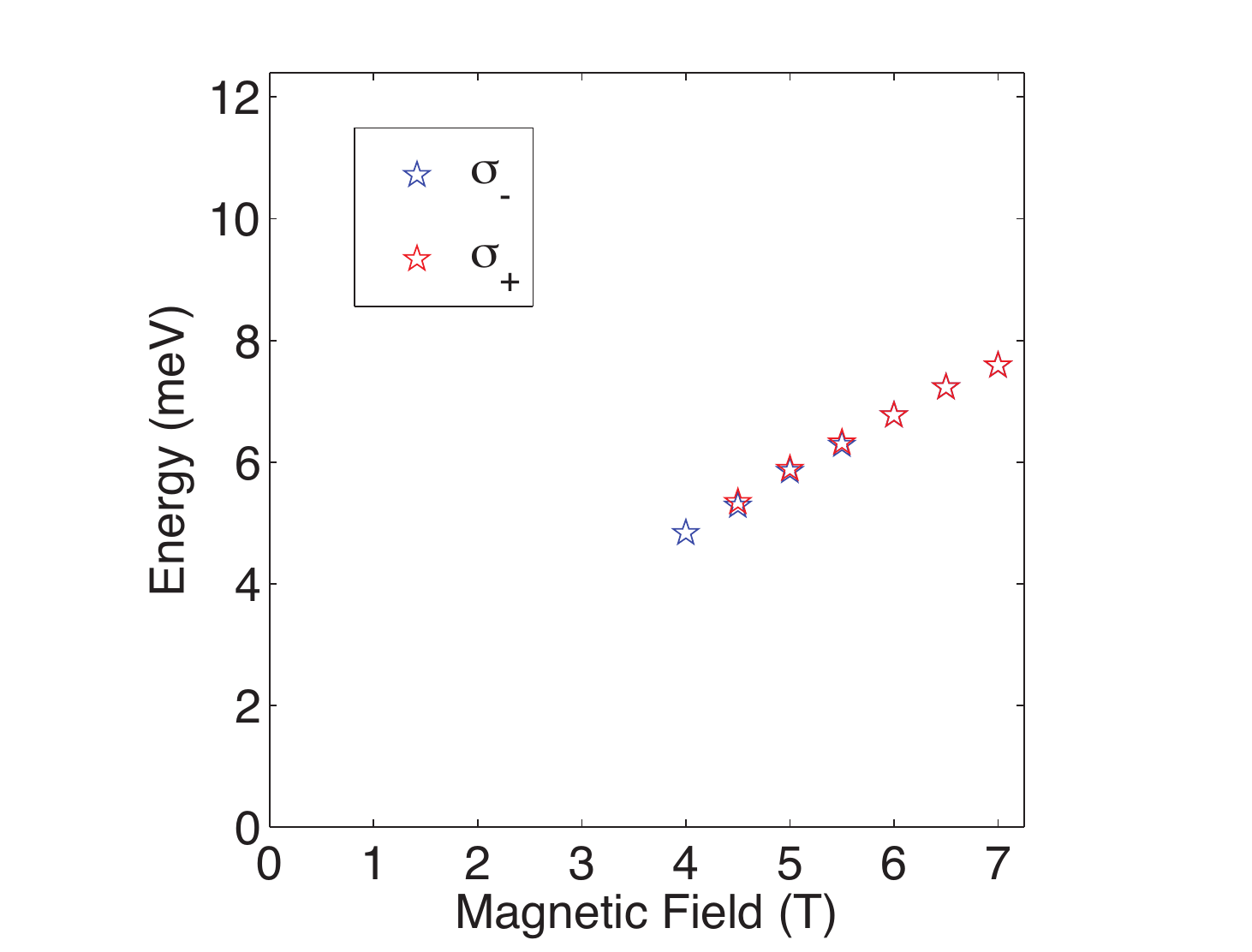}
\caption{\label{figS:smallpeak} Energy of the small unidentified peak in the spectra of $\sigma_-$ and $\sigma_+$ as a function of magnetic field. For the highest three fields the energies overlap. For comparison this figure is on the same scale as in Figure 3 of the main text.}
\end{figure}

\end{document}